\documentclass[journal]{IEEEtran}

\ifCLASSINFOpdf

\else

\fi

\usepackage{epsfig}
\usepackage{graphicx}
\usepackage{amsmath}
\usepackage{amssymb}
\usepackage{booktabs}
\usepackage{multirow}
\usepackage{times}
\usepackage{setspace}
\usepackage{subfigure}
\usepackage{graphicx}
\usepackage{amsfonts}
\usepackage{comment}
\usepackage{color}
\usepackage{tabularx}
\usepackage{makecell}
\usepackage{enumerate}
\usepackage[T1]{fontenc}
\usepackage{hyperref}
\usepackage{url}
\usepackage{bm}
\usepackage{changepage}
\usepackage{lineno}
\usepackage{cite}
\usepackage[numbers]{natbib}
\usepackage{dsfont}
\usepackage{diagbox}
\usepackage{multirow}
\usepackage{multicol}

\ifCLASSINFOpdf
\else
\fi

\begin{document}
%
\title{Emotion Recognition from Multiple Modalities: Fundamentals and Methodologies}



\author{Sicheng~Zhao,~\IEEEmembership{Senior Member, IEEE},~Guoli~Jia,~Jufeng~Yang,~Guiguang~Ding,~Kurt~Keutzer,~\IEEEmembership{Life~Fellow,~IEEE}
\IEEEcompsocitemizethanks{\IEEEcompsocthanksitem S. Zhao is with the Department of Radiology, Columbia University, New York, USA (e-mail: schzhao@gmail.com).\protect
\IEEEcompsocthanksitem G. Jia and J. Yang (corresponding author) are with the College of Computer Science, Nankai University, China (e-mail: exped1230@gmail.com, yangjufeng@nankai.edu.cn).\protect
\IEEEcompsocthanksitem G. Ding is with the School of Software, Tsinghua University, China (e-mail: dinggg@tsinghua.edu.cn).\protect
\IEEEcompsocthanksitem K. Keutzer is with the Department of Electrical Engineering and Computer Sciences, University of California, Berkeley, USA (e-mail: keutzer@berkeley.edu).
}
}

\markboth{IEEE Signal Processing Magazine}
{Shell Zhao{\textit{et al.}}: Emotion Recognition from Multiple Modalities: A Comprehensive Tutorial}

\maketitle

\IEEEpeerreviewmaketitle
Humans are emotional creatures. Multiple modalities are often involved when we express emotions, whether we do so explicitly (e.g., facial expression, speech) or implicitly (e.g., text, image). Enabling machines to have emotional intelligence, i.e., recognizing, interpreting, processing, and simulating emotions, is becoming increasingly important. In this tutorial, we discuss several key aspects of multi-modal emotion recognition (MER). We begin with a brief introduction on widely used emotion representation models and affective modalities. We then summarize existing emotion annotation strategies and corresponding computational tasks, followed by the description of main challenges in MER. Furthermore, we present some representative approaches on representation learning of each affective modality, feature fusion of different affective modalities, classifier optimization for MER, and domain adaptation for MER. Finally, we outline several real-world applications and discuss some future directions.

\section{Introduction}
\label{sec:Introduction}
Emotion is present everywhere in human daily life and can influence or even determine our judgment and decision making~\cite{kahneman2011thinking}. For example, in marketing, a widely advertised brand can generate a mental representation of a product in the consumers' mind and influence their preference and action; inducing sadness and disgust during a shopping trip would respectively increase and decrease consumers' willingness to pay\footnote{\href{https://msbfile03.usc.edu/digitalmeasures/kdiehl/intellcont/Shopping\%20Interdependencies\%20WP-1.pdf}{https://msbfile03.usc.edu/ShoppingInterdependencies}}. In driving, drivers experiencing strong emotions, such as sadness, anger, agitation, and even happiness, are much more likely to be involved in an accident\footnote{\url{https://www.pnas.org/content/113/10/2636}}. In education \textemdash{} especially current online classes during the COVID-19 epidemic period \textemdash{} students’ emotional experiences and interactions with teachers have a big impact on their learning ability, interest, engagement, and even career choices\footnote{\href{https://npjscilearncommunity.nature.com/posts/18507-emotions-in-classrooms-the-need-to-understand-how-emotions-affect-learning-and-education}{https://npjscilearncommunity.nature.com/posts/18507}}.


\begin{figure}[!t]
\begin{center}
\includegraphics[width=0.98\linewidth]{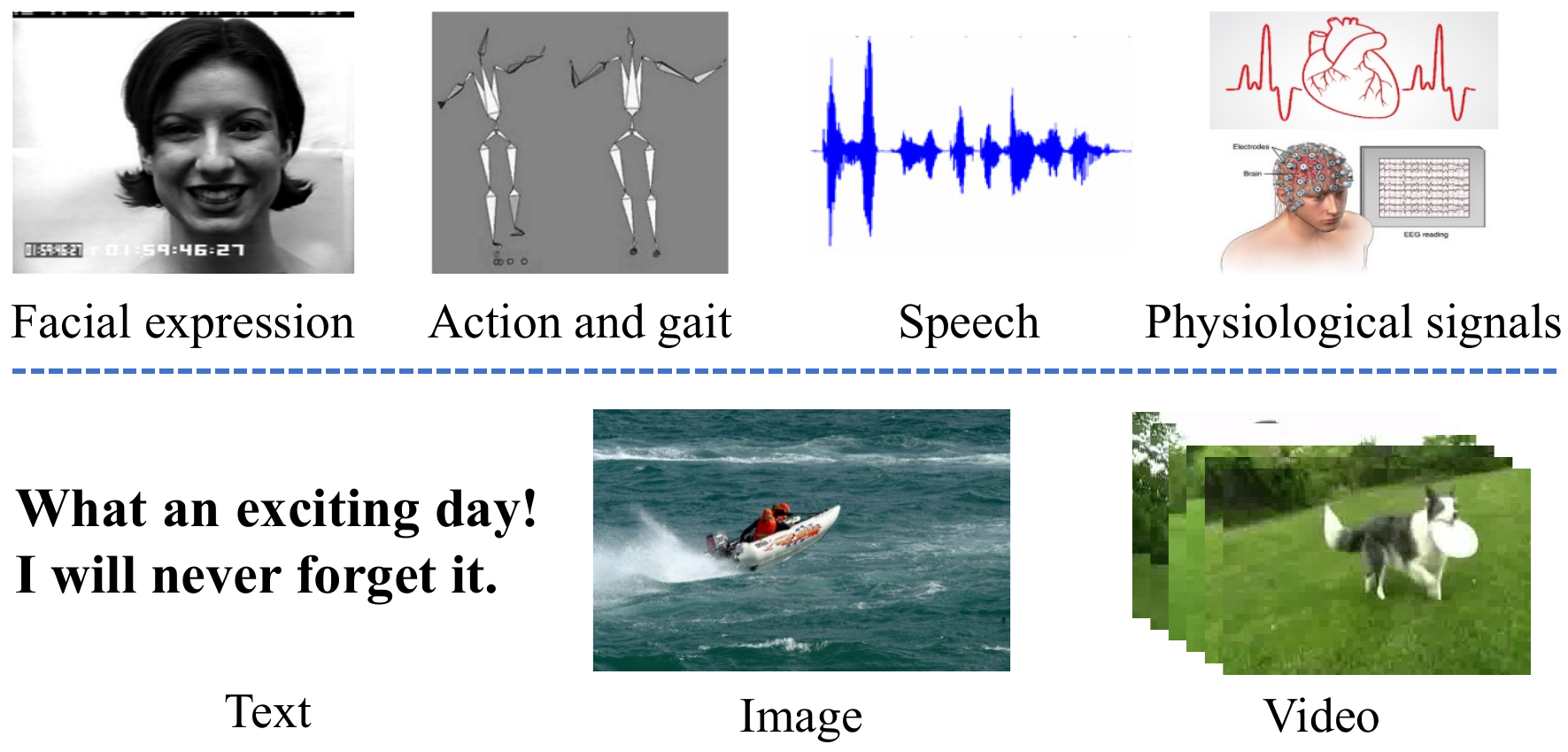}
\caption{Illustration of multiple modalities for emotion recognition: explicit affective cues (top) and implicit affective stimuli (bottom).}
\label{fig:MultimodalExample}
\end{center}
\end{figure}

The importance of emotions in artificial intelligence was recognized decades ago. Minsky, a Turing Award winner in 1970, once claimed that ``\textit{The question is not whether intelligent machines can have any emotions, but whether machines can be intelligent without emotions.}''~\cite{minsky1988society}. Enabling machines to have emotional intelligence i.e., recognizing, interpreting, processing, and simulating emotions has recently become increasingly important with wide potential applications involving human-computer interaction~\cite{schuller2018age}. On the one hand, emotionally intelligent machines can provide more harmonious and personal services for human beings, especially the elderly, disabled, and children. For example, the companion robots that can work with emotions can better meet the psychological and emotional needs of the elderly and help them stay comfortable. On the other hand, by recognizing humans' emotions automatically and in real-time, intelligent machines can better identify humans' abnormal behaviors, send reminders to their relatives and friends, and prevent extreme behaviors to themselves and even to the rest of society. For example, an emotional driving monitoring system can automatically play some soothing music to relax angry drivers who might be dissatisfied with a traffic jam and can remind them to focus on driving safely.

The first step for intelligent machines to express human-like emotions is to recognize and understand humans' emotions typically through two groups of affective modalities: explicit affective cues and implicit affective stimuli. Explicit affective cues correspond to specific physical and psychological changes in humans that can be directly observed and recorded, such as facial expression, eye movement, speech, action, and physiological signals. These signals can be either easily suppressed or masked, or difficult and impractical to capture. Meanwhile, the popularity of mobile devices and social networks enables humans to habitually share their experiences and express their opinions online using text, image, audio, and video. Implicit affective stimuli correspond to these commonly-used digital media, the analysis of which provides an implicit way to infer humans' emotions~\cite{soleymani2017survey}.

Regardless of whether emotions are expressed explicitly or implicitly, there are generally multiple modalities that can contribute to the emotion recognition task, as shown in Fig.~\ref{fig:MultimodalExample}. As compared to uni-modal emotion recognition, multi-modal emotion recognition (MER) has several advantages. First, data complementarity. Cues from different modalities can augment or complement each other. For example, if we see one post from a good friend, "What great weather!", it is of high probability that the friend is expressing a positive emotion; but if there is also an auxiliary image of a storm, we can infer that the text is actually a sarcasm and that a negative emotion is intended to be expressed. Second, model robustness. Due to the influence of many normally occurring factors in data collection, such as sensor device failure, some data modalities might be unavailable, which is especially prevalent in the wild. For example, in the CALLAS dataset containing speech, facial expression, and gesture modalities, the gesture stream is missing for some momentarily motionless users~\cite{wagner2011exploring}. In such cases, the learned MER model can still work with the help of other available modalities. Finally, performance superiority. Joint consideration of the complementary information of different modalities can result in better recognition performance. Meta-analysis indicates that as compared to the best uni-modal counterparts, MER achieves 9.83\% performance improvement on average~\cite{d2015review}.

In this article, we will give a comprehensive tutorial on different aspects of MER, including psychological models, affective modalities, data collections and emotion annotations, computational tasks, challenges, computational methodologies, applications, and future directions. There have been several reviews/surveys on MER related topics~\cite{d2015review,soleymani2017survey,ramachandram2017deep,d2018multimodal,baltruvsaitis2018multimodal}. In particular: \cite{ramachandram2017deep} and \cite{baltruvsaitis2018multimodal} cover different aspects of general multi-modal machine learning with few efforts on emotion recognition; \cite{d2015review} focuses on the
quantitative review and meta-analysis of existing MER systems; \cite{soleymani2017survey} and \cite{d2018multimodal} are survey-style MER articles with the technical emphasis on multi-modal fusion. Differently, this tutorial-style article aims to give a quick and comprehensive MER introduction that is also suitable for non-specialists.

\section{Psychological models}


In psychology, categorical emotion states (CES) and dimensional emotion space (DES) are two representative types of models to measure emotion~\cite{zhao2021affective}. CES models define emotions as a few basic categories, such as binary sentiment (positive and negative, sometimes including neutral), Ekman's six basic emotions (\textit{positive} happiness, surprise and \textit{negative} anger, disgust, fear, sadness), Mikels's eight emotions (\textit{positive} amusement, awe, contentment, excitement, and \textit{negative} anger, disgust, fear, sadness), Plutchik's emotion wheel (eight basic emotion categories by three intensities), and Parrott's tree hierarchical grouping (primary, secondary and tertiary categories). The development of psychological theories motivates CES to be increasingly diverse and fine-grained. DES models employ continuous 2D, 3D, or higher dimensional Cartesian spaces to represent emotions; the most widely used DES model is valence-arousal-dominance (VAD), where valence, arousal, and dominance represent the pleasantness, intensity, and control degree of emotion, respectively.

CES models agree better with humans' intuition, but no consensus has been reached by psychologists on how many discrete emotion categories should be included. Further, emotion is complex and subtle, which cannot be well reflected by limited discrete categories. DES models can theoretically measure all emotions as different coordinate points in the continuous Cartesian space, but the absolute continuous values are beyond users' understanding. These two types of definitions of emotions are related, with possible transformation from CES to DES. For example, anger relates to negative valence, high arousal, and high dominance.

Besides emotion, there are several other widely used concepts in affective computing, such as mood, affect, and sentiment. Emotions can be expected, induced, or perceived. We do not aim distinguishing them in this article. Please refer to~\cite{munezero2014they} for more details on the differences or correlations between these concepts.

\section{Affective Modalities}
In the area of MER, multiple modalities are employed to recognize and predict human emotions. The affective modalities in MER can be roughly divided into two groups based on whether emotions are recognized from humans' physical body changes or from external digital media: explicit affective cues and implicit affective stimuli.
The former group includes facial expression, eye movement, speech, action, gait, and electroencephalogram, all of which can be directly observed, recorded, or collected from an individual.
Meanwhile, the latter group indicates the commonly-used digital media types such as text, audio, image, and video.
We use these data types to store information and knowledge as well as transfer them between digital devices.
In this way, emotions may be implicitly involved and evoked.
Although the efficacy of one specific modality as a reliable channel to express emotions cannot be guaranteed, jointly considering multiple modalities would significantly improve the reliability and robustness~\cite{zeng2009survey}.

\subsection{Explicit Affective Cues}
A \textbf{facial expression} is an isolated motion of one or more human face regions/units, or a combination of such motions.
%
It is commonly agreed that facial expressions can carry informative affective cues and are recognized as one of the most natural and powerful signals to convey the emotional states and intentions of human~\cite{zeng2009survey}.
Facial expression is also a form of nonverbal communication conveying social information between humans.
We can deduce how an individual is feeling by observing his/her \textbf{eyes movement}\footnote{\label{footnote:eyemovement}\url{https://www.frontiersin.org/articles/10.3389/fpsyg.2013.00736/full}}.
The eyes are often viewed as important cues of emotions.
For example, if a person is nervous or lying, the blinking rate of his/her eyes may become slower than normal\textsuperscript{\ref{footnote:eyemovement}}.
Eyes movement signals can be easily collected via an eye tracker system, and have been widely used in human-computer interaction research.
\textbf{Speech} is a significant vocal modal to carry emotions~\cite{zhang2017advanced,akccay2020speech}.
Speakers may express their intentions like asking or declaring by using various intonations, degrees of loudness, and tempo. %
Specifically, emotions can be revealed when people talk with each other, or just mutter to themselves.
As an important part of human body language, \textbf{action}, also conveys massive emotional information.
For instance, an air punch is an act of thrusting one’s clenched fist up into the air, typically as a gesture of triumph or elation.
Similar to action, emotions can be perceived from a person’s \textbf{gait}, i.e., their walking style.
Psychology literature has proven that participants can identify the emotions of a subject by observing the subject's posture, including long strides, collapsed upper body, etc.\footnote{\url{https://www.pnas.org/content/102/45/16518.short}}
Body movement (e.g., walking speed) also plays an important role in the perception of different emotions.
High arousal emotions such as anger and excitement are more associated with rapid movements than low arousal emotions, such as sadness and contentment.
Last but not least, electroencephalogram (\textbf{EEG}), as one representative psychological signal, is another important method to record the electrical and emotional activity of the brain~\cite{subramanian2018ascertain}.
Compared to other aforementioned explicit cues, the collection of EEG signals is typically more difficult and unnatural, regardless of whether electrodes are placed noninvasively along the scalp, or invasively using electrocorticography.
\subsection{Implicit Affective Stimuli}
\textbf{Text} is a form to record the natural language of human beings, which can implicitly carry informative emotions~\cite{giachanou2016like,rahman2020integrating}.
Text has different levels of linguistic components, including word, sentence, paragraph, and article, which are well studied; many off-the-shelf algorithms have been developed to segment text into small pieces.
Then, the affective attribute of each linguistic piece is recognized with the help of a publicly available dictionary like SentiWordNet, and the emotion evoked by the text can be deduced.
A digital \textbf{audio} signal is a representation of sound, typically stored and transferred using a series of binary numbers~\cite{zeng2009survey}.
Audio signals may be synthesized directly or may originate at a transducer such as a microphone or a musical instrument.
Different from speech that mainly focuses on human vocal information and whose content may be translated into natural language, audio is more general including any sound like music or birdsong.
%
An \textbf{image} is a distribution of
colored dots over space\footnote{\url{https://en.wikipedia.org/wiki/Image\#cite_note-1}}.
It is well known that ``a picture is worth a thousand words''.
It has been demonstrated in psychology that emotions can be evoked in humans by images~\cite{joshi2011aesthetics}. The explosive growth of images shared online and the powerful descriptive ability of scenes enable images to become crucial affective stimuli with extensive research efforts attracted~\cite{zhao2021affective}.
%
\textbf{Video} naturally contains multiple modalities at the same time, such as visual, audio, and textual information~\cite{wang2015video}.
That means temporal, spatial, and multi-channel representations can be learned and utilized to recognize the emotions in videos.

\section{Data Collections and Emotion Annotations}
Two steps are usually involved in constructing an MER dataset: data collection and emotion annotation. The collected data can be roughly divided into two categories: selecting from existing data and new recording in specific environments. On the one hand, some data is selected from movies, reviews, videos, and TV shows in online social networks, such as YouTube and Weibo. For example, the review videos in ICT-MMMO and MOUD are collected from YouTube; audio-visual clips are extracted from the TV series in MELD; online reviews from the Food and Restaurants categories are crawled in Yelp; the video-blogs or vlogs typically with one speaker looking at the camera from YouTube are collected in CMU-MOSI to capture the speakers' information. Some collected data provides a transcription of speech either manually (e.g., CMU-MOSI, CH-SMIS) or automatically (e.g., ICT-MMMO, MELD). On the other hand, some data is newly recorded with different sensors in specifically designed environments. For example, the participants’ physiological signals and frontal facial changes induced by music videos are recorded in DEAP.

There are different kinds of emotion annotation strategies. Some datasets have target emotions and do not need to be annotated. For example, in EMODB, each sentence performed by actors corresponds to a target emotion. For some datasets, the emotion annotations are obtained automatically. For example, in Multi-ZOL, the integer sentiment score for each review, ranging from 1 to 10, is regarded as the sentiment label. Several workers are employed to annotate the emotions in some datasets, such as VideoEmotion-8. The datasets with recorded data are usually annotated by participants' self-reporting, such as MAHNOB-HCI. Besides, the emotion labels of most datasets are obtained by major voting. For DES model, `FeelTrace' and `SAM' are often used for annotation. The former one is based on activation-evaluation space, which allows observers to track the emotional content of stimulus as they perceive it over time. The latter one is a tool that accomplishes emotion rating based on different Likert scales. Some commonly used datasets are summarized in Table~\ref{tab:datasets}.

\begin{table*}[!t]
\centering
\caption{Brief summary of released datasets for MER.}
\resizebox{1.0\textwidth}{!}{%
\begin{tabular}{cccccc}
\toprule
dataset  & modalities   & samples  & data sources   & emotion labels  & website  \\
\midrule
IEMOCAP   & face, speech, t-text, video   & 10039 turns  & recording   & ang, sad, hap, dis, fea,   sur, fru, exc, neu & \href{https://sail.usc.edu/iemocap}{link} \\
& & & & VAD on 5 point ratings & \\
YouTube   & face, eye, speech, t-text, video  & 47 videos  & YouTube & pos, neg, neu & \href{http://multicomp.cs.cmu.edu/rsources/youtube-dataset-2}{link} \\
MOUD   & face, speech, t-text, video   & 412 utterances   & YouTube    & pos, neg           & \href{http://web.eecs.umich.edu/$\sim$mihalcea/downloads.html\#MOUD}{link}   \\
ICT-MMMO  & face, eye, speech, t-text, video   & 370 segments   & Youtube, ExpoTV   & pos, neg  & \href{http://multicomp.cs.cmu.edu/resources/ict-mmmo-dataset}{link}  \\
News   Rover    & face, speech, t-text, video  & 929 videos   & News    & pos, neg, neu  & \href{https://www.ee.columbia.edu/n/dvmm/newsrover/sentimentdataset}{link} \\
CMU-MOSI   & face, eye, speech, t-text, video  & 2199 clips   & YouTube    & -3 to 3 sentiment score   & \href{http://multicomp.cs.cmu.edu/resources/cmu-mosi-dataset}{link}  \\
CMU-MOSEI   & face, eye, speech, t-text, video   & 23453 sentences  & YouTube  & hap, sad, ang, fea, dis, sur; -3 to 3 sentiment score   & \href{http://multicomp.cs.cmu.edu/resources/cmu-mosei-dataset}{link}   \\
MELD  & face, speech, t-text, video  & 13708 utterences & TV series Friends  & hap, sad, ang, fea, dis,   sur, neu, non-neu; pos, neg, neu & \href{https://affective-meld.github.io}{link} \\
CH-SIMS  & face, eye, speech, t-text, video  & 2281 segments    & movies, TV series & -1 to 1 sentiment score    & \href{https://github.com/thuiar/MMSA}{link}  \\
 &    &  & variety shows &  &    \\
eNTERFACE'05  & face, speech, video    & 1166 sequences  & recording   & ang, fea, hap, sad, sur   & \href{http://www.enterface.net/enterface05}{link}   \\
SEMAINE  & face, speech, t-text, video   & 959 conversations   & recording     & val, act, pow, exp, int; bas-em, eps, ipa, val   & \href{http://semaine-db.eu}{link}  \\
\midrule
EMDB    & video, SCL, HR   & 52 clips   & films    & ero, hor, neg, pos, sce,   obm; VAD on 9 point ratings   & \href{EMDB@psi.uminho.pt}{link}   \\
DEAP   & face, EEG, GSR, RA, ST & 1280 samples   & recording   & VAD-L on 9 point ratings; F on 5 point ratings    & \href{http://www.eecs.qmul.ac.uk/mmv/datasets/deap}{link}   \\
        & ECG,   BVP, EMG, EOG &     &      &       &     \\
MAHNOB-HCI  & face, eye, audio, EEG  & 532 samples    & recording   & sad, joy, dis, neu, hap,   amu, ang, fea, sur, anx    & \href{https://mahnob-db.eu/hci-tagging}{link} \\
& ECG, GSR, ST,   RA   &    &    & VAD-P on 9 point ratings    &  \\
\midrule
Multi-ZOL  & image, text   & 28K aspect-review pairs   & ZOL  & 0 to 10 sentiment score      & \href{https://github.com/xunan0812/MIMN}{link}  \\
Yelp   & image, text  & 244K images, 44K reviews & Yelp   & sentiment score on 5 point ratings   & \href{https://github.com/PreferredAI/vista-net}{link}   \\
Tourism & image, text   & 1796 weibos & WeiBo   & pos, neg, neu  & \href{https://github.com/wlj961012/Multi-Modal-Event-awareNetwork-for-SentimentAnalysis-in-Tourism}{link}   \\
\midrule
LIRIS-ACCEDE & video (audio, image)   & 9800 clips  & movies   & rank along Valence    & \href{https://liris-accede.ec-lyon.fr}{link}   \\
VideoEmotion-8   & video (audio, image)   & 1101 videos   & YouTube, Flickr     & ang, ant, dis, fea, joy, sad, sur, tru    & \href{http://www.yugangjiang.info/research/VideoEmotions/index.html}{link} \\
Ekman-6  & video (audio, image)  & 1637 videos     & YouTube, Flickr   & ang, dis, fea, joy, sad, sur   & \href{http://bigvid.fudan.edu.cn/data/Ekman.zip}{link}   \\
\bottomrule
\end{tabular}
}
\label{tab:datasets}
Modalities: t-text, EEG, PPS, GSR, RA, ST, ECG, BVP, EMG, EOG, SCL, and HR are short for transcript text, Electroencephalogram, Peripheral-Physiological-Signal, Galvanic-Skin-Response, Respiration-Rmplitude, Skin Temperature, Electrocardiogram, Blood-Volume-Pressure, Electromyogram, Electrooculogram, Skin-Conductance-Level, and Heart-Rate, respectively. Emotion labels: ang, sad, hap, dis, fea, sur, fru, exc, neu, pos, neg, joy, amu, anx, ero, hor, sce, obm, ant, and tru are short for angry, sadness, happiness, disgust, fear, surprise, frustration, excited, neutral, positive, negative, joy, amusement, anxiety, erotic, horror, scenery, object-manipulation, anticipation, and trust, respectively; -L, F, and -P are short for liking, familiarity, and predictability, respectively.
\end{table*}

\section{Computational Tasks}

Given multi-modal affective signals, we can conduct different MER tasks, including classification, regression, detection, and retrieval. In this section, we will briefly introduce what these tasks do.

\subsection{Emotion Classification}
In the emotion classification task, we assume that one instance can only belong to one or a fixed number of emotion categories, and the goal is to discover class boundaries or class distributions in the data space~\cite{giachanou2016like}.
Current works mainly focus on the manual design of multi-modal features and classifiers or employing deep neural networks in an end-to-end manner.
As defined as a single label learning (SLL) problem, MER assigns a single dominant emotion label to each sample.
However, emotion may be a mixture of all components from different regions or sequences rather than a single representative
emotion.
Meanwhile, different people may have different emotional reactions to the same stimulus, which is
caused by a variety of elements like  personality.
Thus, multi-label learning (MLL) has been utilized to study the problem where one instance is associated with multiple emotion labels.
Recently, to address the problem that MLL does not fit some real applications well where the overall distribution of different labels' importance matters, label distribution learning (LDL) is proposed to cover a certain number of labels, representing the degree to which each emotion label describes the instance~\cite{yang2017aaai}.

\subsection{Emotion Regression}
Emotion regression aims to learn a mapping function that can effectively associate one instance with continuous emotion values in a Cartesian space.
%
The most common regression algorithms for MER aim to assign the average dimension values to the source data.
%
To deal with the inherent subjectivity characteristic of emotions, researchers propose to predict the continuous probability distribution of emotions which are represented in dimensional valence-arousal (VA) space.
Specifically, VA emotion labels can be represented by a Gaussian mixture model (GMM), and then the emotion distribution prediction can be formalized as a parameter learning problem~\cite{zhao2017continuous}.

\subsection{Emotion Detection}
As the raw data does not ensure carrying emotions, or only part of the data can evoke emotional reactions, emotion detection aims to find out which kind of emotion lies where in the source data.
For example, a restaurant review on Yelp might be ``This location is conveniently located across the street from where I work, being walkable is a huge plus for me! Food wise, it's the same as almost every location I've visited so there's nothing much to say there. I do have to say that the customer service is hit or miss.''
Meanwhile, the overall rating score is three stars out of five. This review contains different emotions and attitudes: positive in the first sentence, neutral in the second sentence, and negative in the last sentence.
As such, it is crucial for the system to detect which sentence corresponds to which emotion.
Another example is affective region detection in images~\cite{she2020wscnet}.

\subsection{Emotion Retrieval}
How to search affective content based on human perception is another meaningful task.
%
The existing framework first detects local interest patches or sequences in the query and candidate data sources.
Then, it discovers all matched pairs by determining whether the distance between two patches or sequences is less
than a given fixed threshold.
The similarity score between the query and each candidate is calculated
as the quantity of matched components, followed by ranking
the candidates of this query accordingly.
While an affective retrieval system is useful for obtaining online content with desired emotions from a massive repository~\cite{zhao2021affective}, again the abstract and subjective characteristics make the task challenging and difficult to evaluate.

\section{Challenges}

As stated in Section~\ref{sec:Introduction}, multi-modal emotion recognition (MER) has several advantages as compared to uni-model emotion recognition but it also faces more challenges.

\subsection{Affective Gap}
The affective gap is one main challenge for MER, which measures the inconsistency between extracted features and perceived high-level emotions. The affective gap is even more challenging than the semantic gap in objective multimedia analysis. Even if the semantic gap is bridged, there might still exist an affective gap. For example, a blooming rose and a faded rose both contain a rose but can evoke different emotions. For the same sentence, different voice intonations may correspond to totally different emotions. Extracting discriminative high-level features and especially those related to emotions can help to bridge the affective gap. The main difficulty lies in how to evaluate whether the extracted features are related to emotions.

\subsection{Perception Subjectivity}
Due to many personal, contextual, and psychological factors, such as the cultural background, personality, and social context, different people might have different emotional responses to the same stimuli~\cite{zhao2021affective}. Even if the emotion is the same, their physical and psychological changes can also be quite different. For example, all the 36 videos in the ASCERTAIN dataset for MER are labeled with at least four out of seven different valence and arousal scales by 58 subjects~\cite{subramanian2018ascertain}. This clearly indicates that some subjects have the opposite emotional reactions to the same stimuli. Take a short video with storm and thunder for instance, some people may feel in awe because they have never seen such extreme weather, some may feel fear because of the loud thunder noise, some may feel excited to capture such rare scenes, some may feel sad because they have to cancel their travel plans, etc. Even for the same emotion (e.g., excitement), there are different reactions, such as facial expression, gait, action, and speech. For the subjectivity challenge, one direct solution is to learn personalized MER models for each subject. From the perspective of stimuli, we can also predict the emotion distribution when a certain number of subjects are involved. Besides the content of the stimuli and direct physical and psychological changes, jointly modeling the above-mentioned personal, contextual, and psychological factors would also contribute to the MER task.

\subsection{Data Incompleteness}
Because of the presence of many inevitable factors in data collection, such as sensor device failure, the information in specific modalities might be corrupted, which results in missing or incomplete data. Data incompleteness is a common phenomenon in real-world MER tasks. For example, for explicit affective cues, an EEG headset might record contaminated signals or even fail to record any signal; at night, the cameras cannot capture clear facial expressions. For implicit affective stimuli, one user might post a tweet only containing an image (without text); for some videos, the audio channel does not change much. In such cases, the simplest feature fusion method, i.e., early fusion, does not work, because we cannot extract any features given no captured signal. Designing effective fusion methods that can deal with data incompleteness is a widely employed strategy.

\subsection{Cross-modality Inconsistency}
Different modalities of the same sample may conflict with each other and thus express different emotions. For example, facial expression and speech can be easily suppressed or masked to avoid being detected, but EEG signals that are controlled by the central nervous systems can reflect humans' unconscious body changes. When people post tweets on social media, it is very common that the images are not semantically correlated to the text. In such cases, an effective MER method is expected to automatically evaluate which modalities are more reliable, such as by assigning a weight to each modality.

\subsection{Cross-modality Imbalance}
In some MER applications, different modalities may contribute unequally to the evoked emotion.
For example, online news plays an important role in our daily lives, and in addition to understanding the preferences of readers, predicting their emotional reactions is of great value in various applications, such as personalized advertising.
However, a piece of online news usually includes imbalanced texts and images, i.e., the length of the article may be very long with lots of detailed information, while only one or two illustrations are inserted into the news.
Potentially more problematic, the editor of the news may select a neutral image for an article with an obvious sentiment.

\begin{figure*}[!t]
\begin{center}
\includegraphics[width=0.98\linewidth]{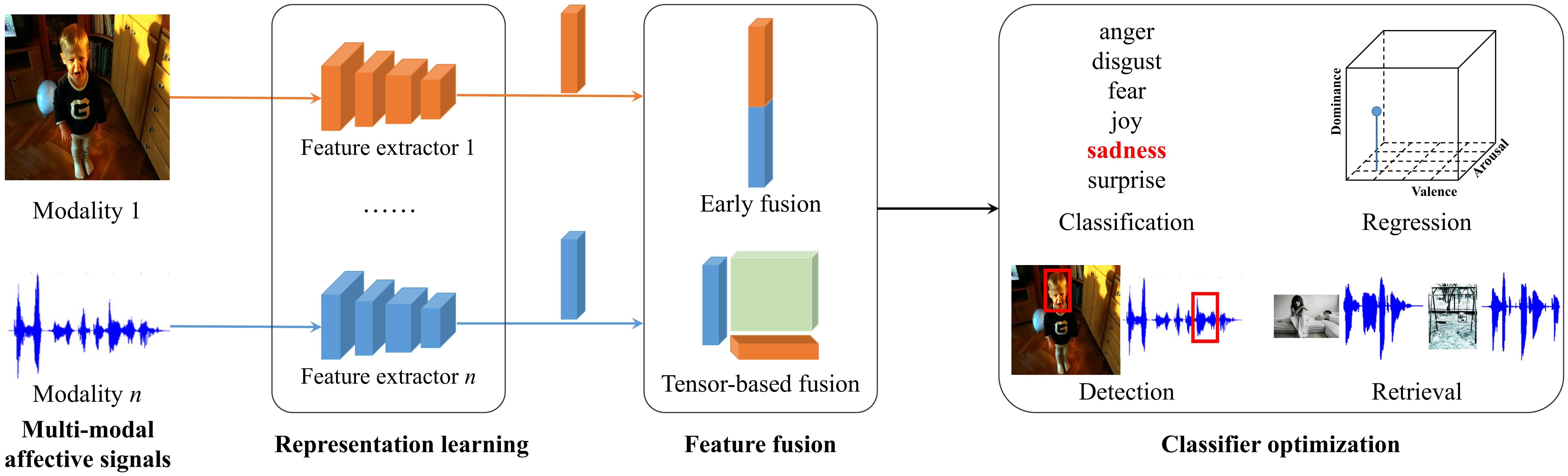}
\caption{Illustration of a widely used MER framework, which consists of three components: representation learning to extract feature representations, feature fusion to combine features from different modalities, and classifier optimization to learn specific task models (e.g., classification, regression, detection, and retrieval). $n$ is the number of different modalities.}
\label{fig:MERFramework}
\end{center}
\end{figure*}

\subsection{Label Noise And Absence}
Existing MER methods, especially the ones based on deep learning, require large-scale labeled data for training. However, in real-world applications, labeling the emotions in the ground-truth generation is not only prohibitively expensive and time-consuming but also highly inconsistent, which results in a large amount of data but with few or even no emotion labels. With the increasingly diverse and fine-grained emotion requirement, we might have enough training data for some emotion categories but not for others. One alternate solution to manual annotation is to leverage the tags or keywords of social tweets as emotion labels, but such labels are incomplete and noisy. As such, designing effective algorithms for unsupervised/weakly-supervised learning and few/zero shot learning can provide potential solutions.

Meanwhile, we might have sufficient labeled affective data in one domain, such as synthetic facial expression and speech. The problem turns to how to effectively transfer the trained MER model on the labeled source domain to another unlabeled target domain. The presence of domain shift causes significant performance decay when a direct transfer is used~\cite{zhao2020review}. Multi-modal domain adaptation and domain generalization can help to mitigate such domain gap.
Practical settings, such as multiple source domains, should also be considered.

\section{Computational Methodologies}

Generally, there are three components in an MER framework with sufficient labeled training data in the target domain: representation learning, feature fusion, and classifier optimization, as shown in Fig.~\ref{fig:MERFramework}. In this section, we will introduce these components. Further, we will describe domain adaptation when there is no labeled training data in the target domain and when sufficient labeled data is available in another related source domain.

\subsection{Representation Learning of Each Affective Modality}
To represent the text as a form that can be understood by computers, the following aspects are required: first, representing the symbolic words as real numbers for the next computation; second, modeling the semantic relationships; and finally, obtaining a unified representation for the whole text~\cite{giachanou2016like}.
In the beginning, words are represented by one-hot vectors with the length of vocabulary size, where for the $t$-th word in the vocabulary, $w_t$, only the position $t$ is $1$ and the other positions are $0$.
As the scale of the data increases, the dimension of this one-hot vector increases dramatically.
Later, researchers use language models to train word vectors by predicting context, obtaining word vectors with vectors of fixed dimension.
Popular word vector representation models include word2vec, GLOVE, BERT, XLNet, and so on.
The text feature extraction methods have developed from simple ones to complex ones as well.
Text features can be obtained by simply averaging word vectors.
A recurrent neural network (RNN) is used to model the sequential relations of words in the text.
A convolutional neural network (CNN) which has been widely used in the computer vision community, is also used to extract contextual relations between words.
To date, plenty of methods have been developed to design representative features for emotion stimuli in audios~\cite{zhang2017advanced,akccay2020speech}.
It has been found that audio features such as pitch, log energy, zero-crossing rate, spectral features, voice quality, and jitter, are useful in emotion recognition. The ComParE acoustic feature set is commonly used as the baseline set for the ongoing Computation Paralinguistics Challenge series since 2013.
However, because of possible high similarities in certain emotions, single type of audio feature is not discriminative enough to classify emotions.
To solve this problem, some approaches propose to combine different types of features.
Recently, with the development of deep learning, CNN is shown to achieve state-of-the-art performance on large-scale tasks in many domains dealing with natural data, and audio emotion recognition is of course also included.
Audio is typically transferred into a graphical representation, such as a spectrogram, to be fed into a CNN. Since CNN uses shared weight filters and pooling to give the model better spectral and temporal invariant properties, it typically yields better generalized and more robust models for emotion recognition.
Researchers have designed informative representations for emotional stimuli in images.
In general, images can be divided into two types, \textit{non-restrictive images} and \textit{facial expression images}.
For the former, e.g., natural images, various hand-crafted features including color, texture, shape, composition, etc., are developed to represent image emotion in the early years~\cite{zhao2021affective}.
These low-level features are developed with inspiration from psychology and art theory.
Later, mid-level features based on the visual concepts are presented to bridge the gap between the pixels in images and the emotion labels.
The most representative engine is SentiBank, which is composed of $1,200$ adjective-noun pairs and shows remarkable and robust recognition performance among all the hand-engineering features.
In the era of deep learning, CNN is regarded as a strong feature extractor in an end-to-end manner.
Specifically, to integrate various representations of different levels, features are extracted from multiple layers of CNN.
Meanwhile, an attention mechanism is employed to learn better emotional representations of specific local affective regions~\cite{she2020wscnet}.
For the facial expression images, firstly the human face is detected and aligned, and then the face landmarks are encoded for the recognition task.
Note that for those non-restrictive images that contain human faces by chance, facial expression can be treated as an important mid-level cue.
In the above, we have mentioned how to identify emotions in the isolated modalities.
Here, we first focus on perceiving emotions from successive frames.
Then, we introduce how to build joint representation for videos.
Compared to a single image, a video contains a series of images with temporal information~\cite{wang2015video}.
To build representations of videos, a wide range of methods has been proposed.
Early methods mainly utilize hand-crafted local representations in this field, which include color, motion, and shot cut rate.
With the advent of deep learning, recent methods extract discriminative representations by adopting a 3D CNN that captures the temporal information encoded in multiple adjacent frames.
After extracting modality-specific features in videos, integrating different types of features could obtain more promising results and improve the performance.
To perceive emotions, there are mainly two aspects of ways to learn the representations of gait~\cite{bhattacharya2020step}.
For one thing, we can explicitly model the posture and movement information that is related to the emotions.
To model this information, we first extract the skeletal structure of a person and then represent each joint of the human body using the 3D coordinate system.
After getting these coordinates, the angles, distance, or area among different joints (posture information), velocity/acceleration (movement information), their co-variance descriptors, etc. can be easily extracted.
%
%
For another thing, high-level emotional representations can be modeled from gait by long short-term memory (LSTM), deep convolutional neural networks, or graph convolutional networks.
Some methods extract optical flow from gait videos and then extract sequence representations using these networks.
Other methods learn skeletal structures of the gait and then feed them into multiple networks to extract discriminate representations.
Since various information about emotions, such as frequency band, electrodeposition, and temporal information, can be explored from the brain's response to emotional stimuli, EEG signals are widely used in emotion analysis~\cite{subramanian2018ascertain}.
To extract discriminative features for EEG emotion recognition, differential entropy features from frequency band or electrodeposition relationship are very popular in previous work.
Besides hand-crafted features, we can also directly apply end-to-end deep learning neural networks such as CNN and RNN on the raw EEG signals to obtain powerful deep features~\cite{liu2018realtime}.
Inspired by the learning pattern of humans, spatial-wise attention mechanisms are successfully applied to extract more discriminative spatial information.
Furthermore, considering that EEG signals contain multiple channels, a channel-wise attention mechanism can also be integrated into CNN to exploit the inter-channel relationship among feature maps.

\subsection{Feature Fusion of Different Affective Modalities}
Feature fusion, as one key research topic in MER, aims to integrate the representations from multiple modalities to predict either a specific category or a continuous value of emotions. Generally, there are two strategies: model-free fusion and model-based fusion~\cite{ramachandram2017deep,baltruvsaitis2018multimodal}.

\begin{figure}[!t]
\begin{center}
\includegraphics[width=0.95\linewidth]{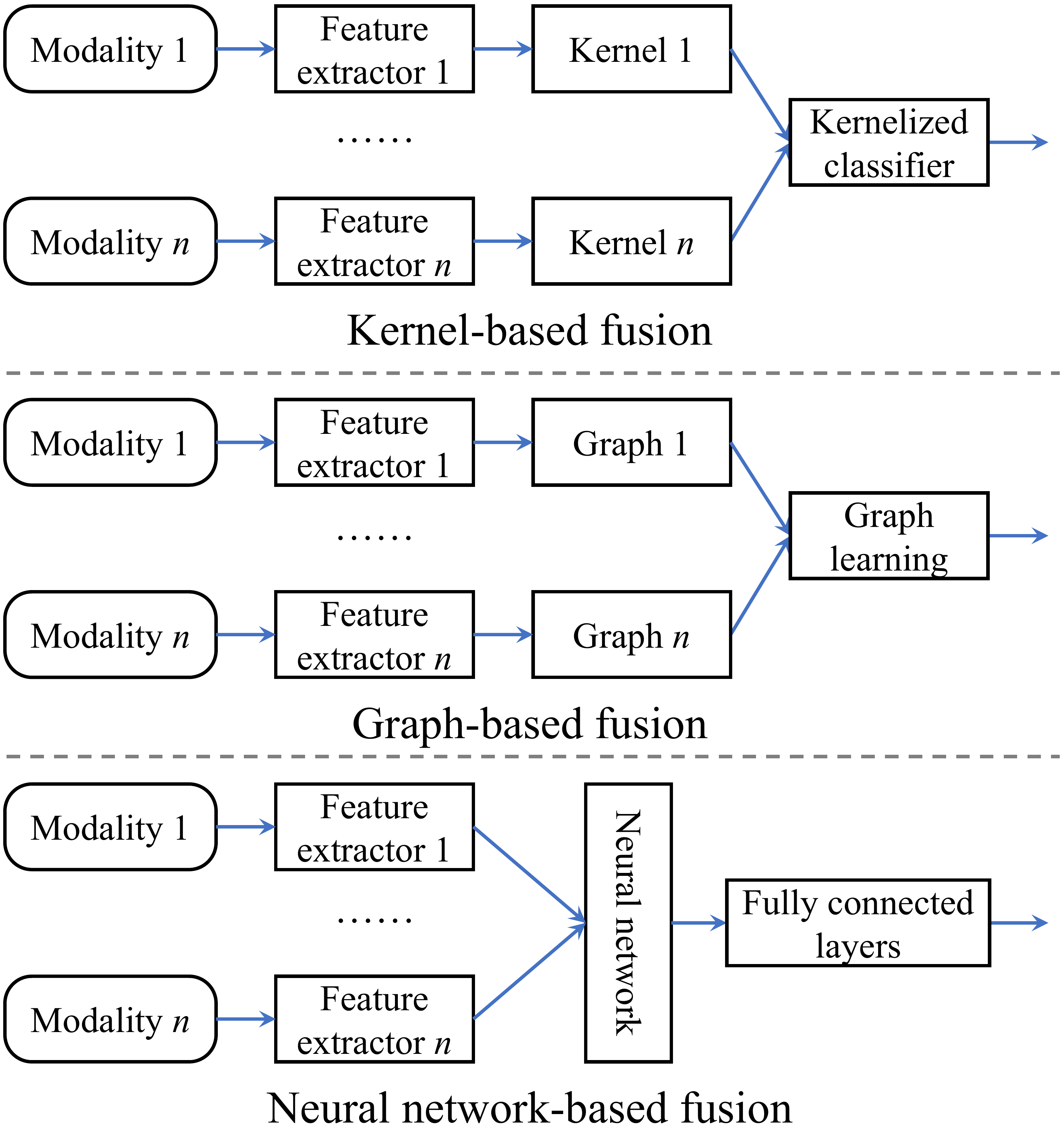}
\caption{Illustration of different model-based fusion strategies, where $n$ is the number of different modalities.}
\label{fig:fusion}
\end{center}
\end{figure}

\textbf{Model-free fusion} that is not directly dependent on specific learning algorithms has been widely used for decades. We can divide it into early fusion, late fusion, and hybrid fusion~\cite{wagner2011exploring}. All these fusion methods can be extended from existing uni-modal emotion recognition classifiers. \textit{Early fusion}, also named \textit{feature-level fusion}, directly concatenates the feature representations from different modalities as a single representation. It is the most intuitive method to fuse different representations by exploiting the interactions between different modalities at an early stage and only requires training a single model. But since the representations from different modalities might significantly differ, we have to consider the time synchronization problem to transform these representations into the same format before fusion. When one or more modalities are missing, such early fusion would fail. \textit{Late fusion}, also named \textit{decision-level fusion}, instead integrates the prediction results from each single modality. Some popular mechanisms include averaging, voting, and signal variance. The advantages of late fusion include (1) flexibility and superiority -- the optimal classifiers can be selected for different modalities;  and (2) robustness -- when some modalities are missing, late fusion can still work. However, the correlations between different modalities before decision are ignored. \textit{Hybrid fusion} combines early fusion and late fusion to exploit their advantages in a unified framework but with higher computational cost.

\textbf{Model-based fusion} that explicitly performs fusion during the construction of the learning models has been paid more attention~\cite{ramachandram2017deep,baltruvsaitis2018multimodal}, as shown in Fig.~\ref{fig:fusion}, since model-free fusion is based on some simple techniques that are not specifically designed for multimodal data. For shallow models, kernel-based fusion and graph-based fusion are two representative methods; for recent popular deep models, neural network-based fusion, attention-based fusion, and tensor-based fusion are often used.

\textit{Kernel-based fusion} is extended based on classifiers that contain kernels, such as SVM. For different modalities, different kernels are used. The flexibility in kernel selection and convexity of the loss functions make multiple kernel learning fusion popular in many applications, including MER. However, during testing, these fusion methods rely on the support vectors in the training data, which results in large memory cost and inefficient reference. \textit{Graph-based fusion} constructs separate graphs or hypergraphs for each modality, combines these graphs into a fused one, and learns the weights of different edges and modalities by graph-based learning. It can well deal with the data incompleteness problem simply by constructing graphs based on available data. Besides the extracted feature representations, we can also incorporate prior human knowledge into the models by corresponding edges. However, the computational cost would increase exponentially when more training samples are available.

\textit{Neural network-based fusion} employs a direct and intuitive strategy to fuse the feature representations or predicted results of different modalities by a neural network. \textit{Attention-based fusion} uses some attention mechanisms to obtain the weighted sum of a set of feature representations with scalar weights that are dynamically learned by an attention module. Different attention mechanisms correspond to fusing different components. For example, spatial image attention measures the importance of different image regions. Image and text co-attention employs symmetric attention mechanisms to generated both attended visual and attended textual representations. Parallel co-attention and alternating co-attention methods can be used to respectively generate attention for different modalities simultaneously and one by one. Recently, a Multimodal Adaptation Gate (MAG) is designed to enable Transformer-based contextual word representations, such as BERT and XLNet, to accept multi-modal nonverbal data~\cite{rahman2020integrating}. Based on the attention conditioned on the nonverbal behaviors, MAG can essentially map the informative multiple modalities to a vector with a trajectory and magnitude. \textit{Tensor-based fusion} tries to exploit the correlations of different representations by some specific tensor operations, such as outer product and polynomial tensor pooling. These fusion methods for deep models are capable of learning from a large amount of data in an end-to-end manner with good performance but suffer from low interpretability.

One important property of the above-mentioned feature fusion methods is whether they support temporal modeling for MER in videos. It is obvious that early fusion can while late fusion and hybrid fusion cannot, since the predicted results based on each modality are already known before late fusion. For model-based fusion, excluding kernel-based fusion, all others can be used for temporal modeling, such as hidden Markov models (HMM) and conditional random fields (CRF) for graph-based fusion methods, and RNN and LSTM networks for neural network-based fusion.

\subsection{Classifier Optimization for Multi-modal Emotion Recognition}
For the text represented as a sequence of word embeddings, the most popular approaches to leverage the semantics among words are RNN and CNN.
LSTM, as a typical RNN, contains a series of cells with the same structure.
Every cell takes a word embedding and the hidden state from the last cell as input, computes the output, and updates the hidden state for the following cell.
The hidden state records the semantics of previous words.
CNN computes local contextual features among consecutive words through convolution operations.
And average pooling or max-pooling layers are used to further integrate the obtained features for the following sentiment classification.
Recently, researchers begin to use Transformer-based methods, e.g., BERT and GPT-3.
Transformer is implemented as a series of modules containing a multi-head self-attention layer followed by a normalization layer, a feed-forward network, and another normalization layer.
The order of words in the text is also represented by another position embedding layer.
Compared with RNN, transformer does not require sequential processing of words, which improves the parallelizability.
And compared with CNN, transformer can model relationships between more distant words.
The classification approaches used in audio emotion recognition generally include the following two options: traditional methods and deep learning-based methods.
For traditional methods, HMM is a representative method because of its capability of capturing dynamic characteristics of sequential data.
SVM is also widely utilized in audio emotion recognition.
Deep learning-based methods have become more and more popular since they are not restricted by the classical independence assumptions of HMM models.
Among these methods, sequence-to-sequence models with attention have shown success in an end-to-end manner.
Recently, some approaches significantly extend the state of the art in this area by developing deep hybrid convolutional and recurrent models~\cite{akccay2020speech}.

\begin{table*}[!t]
\centering
\caption{Quantitative comparison of some representative methods for multi-modal emotion recognition on five widely used datasets using GLOVE as word embeddings.}
\resizebox{1.0\textwidth}{!}{%
\begin{tabular}{c|ccccc|cc|cc|cc|cccccc}
\toprule
dataset          & \multicolumn{5}{c|}{CMU-MOSI}         & \multicolumn{2}{c|}{YouTube}     & \multicolumn{2}{c|}{ICT-MMMO}   & \multicolumn{2}{c|}{MOUD}         & \multicolumn{6}{c}{IEMOCAP}               \\
\midrule
train : val : test & \multicolumn{5}{c|}{1284 : 229 : 686} & \multicolumn{2}{c|}{30 : 5 : 11} & \multicolumn{2}{c|}{11 : 2 : 4} & \multicolumn{2}{c|}{49 : 10 : 20} & \multicolumn{6}{c}{3 : 1 : 1}             \\
\midrule
\diagbox{method}{metric}                 & $\rm A_2$$\uparrow$    & F1$\uparrow$    & $\rm A_7$$\uparrow$    & M$\downarrow$  & C$\uparrow$  & $\rm A_3$$\uparrow$             & F1$\uparrow$             & $\rm A_2$$\uparrow$             & F1$\uparrow$            & $\rm A_2$$\uparrow$              & F1$\uparrow$             & $\rm A_9$$\uparrow$ & F1$\uparrow$ & $\rm M_V$$\downarrow$ & $\rm C_V$$\uparrow$ & $\rm M_A$$\downarrow$ & $\rm C_A$$\uparrow$ \\
\midrule
SVM  & 71.6 & 72.3 & 26.5 & 1.100 & 0.559 & 42.4 & 37.9 & 68.8 & 68.7 & 60.4 & 45.5 & 24.1 & 18.0 & 0.251 & 0.060 & 0.546 & 0.540 \\
RF   & 56.4 & 56.3 & 21.3 & -     & -     & 49.3 & 49.2 & 70.0 & 69.8 & 64.2 & 63.3 & 27.3 & 25.3 & -     & -     & -     & -     \\
THMM & 50.7 & 45.4 & 17.8 & -     & -     & 42.4 & 27.9 & 53.8 & 53.0 & 58.5 & 52.7 & 23.5 & 10.8 & -     & -     & -     & - \\
\midrule
MV-LSTM & 73.9 & 74.0 & 33.2 & 1.019 & 0.601 & 45.8 & 43.3 & 72.5 & 72.3 & 57.6 & 48.2 & 31.3 & 26.7 & 0.257 & 0.020 & 0.513 & 0.620 \\
BC-LSTM & 73.9 & 73.9 & 28.7 & 1.079 & 0.581 & 47.5 & 47.3 & 70.0 & 71.1 & 72.6 & 72.9 & 35.9 & 34.1 & 0.248 & 0.070 & 0.593 & 0.400 \\
TFN     & 74.6 & 74.5 & 28.7 & 1.040 & 0.587 & 47.5 & 41.0 & 72.5 & 72.6 & 63.2 & 61.7 & 36.0 & 34.5 & 0.251 & 0.040 & 0.521 & 0.550 \\
MARN    & 77.1 & 77.0 & 34.7 & 0.968 & 0.625 & 54.2 & 52.9 & 86.3 & 85.9 & 81.1 & 81.2 & 37.0 & 35.9 & 0.242 & 0.100 & 0.497 & 0.650 \\
MFN     & 77.4 & 77.3 & 34.1 & 0.965 & 0.632 & 61.0 & 60.7 & 87.5 & 87.1 & 81.1 & 80.4 & 36.5 & 34.9 & 0.236 & 0.111 & 0.482 & 0.645 \\
\bottomrule
\end{tabular}
}
Evaluation metrics: $\rm A_N$ means emotion classification accuracy where N denotes the number of emotion classes, $\rm A_N$ and F1 are in percentage, M is short for mean absolute error, C indicates the Pearson correlation, and $\rm _V$, $\rm _A$ correspond to the results of valence and arousal (the same for Table~\ref{tab:CMU-MOSI}).
\label{tab:GLOVE}
\end{table*}

\begin{table}[!t]
\centering\small
\caption{Quantitative comparison of some representative methods for multi-modal emotion recognition on the CMU-MOSI dataset using BERT or XLNet as word embeddings.}
\resizebox{1.0\linewidth}{!}{%
\begin{tabular}{c|cccc}
\toprule
\diagbox{method}{metric} & $\rm A_2$$\uparrow$   & F1$\uparrow$   & M$\downarrow$   & C$\uparrow$  \\
\midrule
TFN   & 74.8/78.2 & 74.1/78.2 & 0.955/0.914 & 0.649/0.713 \\
MARN  & 77.7/78.3 & 77.9/78.8 & 0.938/0.921 & 0.691/0.707 \\
MFN   & 78.2/78.3 & 78.1/78.4 & 0.911/0.898 & 0.699/0.713 \\
FT    & 83.5/84.7 & 83.4/84.6 & 0.739/0.676 & 0.782/0.812 \\
MAG   & 84.2/85.7 & 84.1/85.6 & 0.712/0.675 & 0.796/0.821 \\
\midrule
Human & 85.7      & 87.5      & 0.710       & 0.820 \\
\bottomrule
\end{tabular}
}
The numbers on the left side and the right side of “/” are the MER results based on BERT and XLNet, respectively.
\label{tab:CMU-MOSI}
\end{table}

In the early years, similar to this task in other modalities, multiple hand-crafted image features are integrated and input into SVM to train classifiers.
Then, based on deep learning, the classifier and feature extractor are connected and optimized in an end-to-end manner by corresponding loss functions like cross-entropy loss~\cite{hu2018structure}.
Besides, popular metric losses such as triplet loss and N-pair loss also take part in the network optimization to obtain more discriminative features.
With the above learning paradigm, each image is predicted as a single dominant emotion category.
However, based on the theories of psychology, an image may evoke multiple emotions in viewers, which leads to an ambiguous problem.
To address the problem, label distribution learning is employed to predict a concrete relative degree for each emotion category, where Kullback–Leibler divergence is the most popular loss function.
Some informative and attractive regions of an image always determine the emotion of the image.
Therefore, a series of architecture with extra attention or detection branch is constructed.
With the optimization for multiple tasks including attention and original task, a more robust and discriminative model is obtained.
Most existing methods employ a two-stage pipeline to recognize video emotion, i.e., extracting visual and/or audio features and training classifiers.
For training classifiers, many machine learning methods have been investigated to model the mapping between video features and discrete emotion categories, including SVM, GMM, HMM, dynamic Bayesian networks (DBNs), and conditional random fields (CRF).
Despite the above methods have contributed to the development of emotion recognition in videos, recent methods are proposed to recognize video emotions in an end-to-end manner based on deep neural networks due to their superior capability~\cite{kahou2015video}.
CNN-based methods first employ 3D convolutional neural networks to extract high-level spatio-temporal features which contain affective information, and then use fully connected layers to classify emotions.
Finally, the models are followed by the loss function to optimize the whole network.
Inspired by the human process of perceiving emotions, CNN-based methods employ the attention mechanism to emphasize emotionally relevant regions of frames or segments in each video.
Furthermore, considering the polarity-emotion hierarchy constraint, recent methods propose polarity-consistent cross-entropy loss, to guide the attention generation.

The gait of a person can be represented as a sequence of 2D or 3D joint coordinates for each frame in the walking videos.
To leverage the inherent affective cues in the coordinates of joints, many classifiers or architectures have been used to extract affective features in the gait.
LSTM networks contain many special units, i.e., memory cells, and can store the joint coordinate information from particular time steps in a long-time data sequence.
Thus, it is used in some early work of gait emotion recognition.
The hidden features of the LSTM can be further concatenated with the hand-crafted affective features and are then fed into a classifier (e.g., SVM or random forest (RF)) to predict emotions.
Recently, another popular network used in gait emotion prediction is the spatial-temporal graph convolutional network (ST-GCN).
ST-GCN is initially proposed for action recognition from human skeletal graphs.
`Spatial' represents the spatial edges in the skeletal structure, which are the limbs that connect the body joints.
`Temporal' refers to temporal edges, and they connect the positions of each joint across different frames.
ST-GCN can be easily implemented as a spatial convolution followed by a temporal convolution, which is similar to the deep convolutional networks.
EEG-based emotion recognition usually employs various classifiers such as SVM, decision trees, and $k$-nearest neighbor to classify hand-crafted features in the early stage.
Later, since CNN and RNN are good at extracting spatial information and temporal information of EEG signals, respectively, end-to-end structures such as cascade convolutional recurrent network (which combines CNN and RNN), LSTM-RNN, and parallel convolutional recurrent neural networks are successfully designed and applied to emotion recognition tasks.

\begin{figure*}[!t]
\begin{center}
\includegraphics[width=0.98\linewidth]{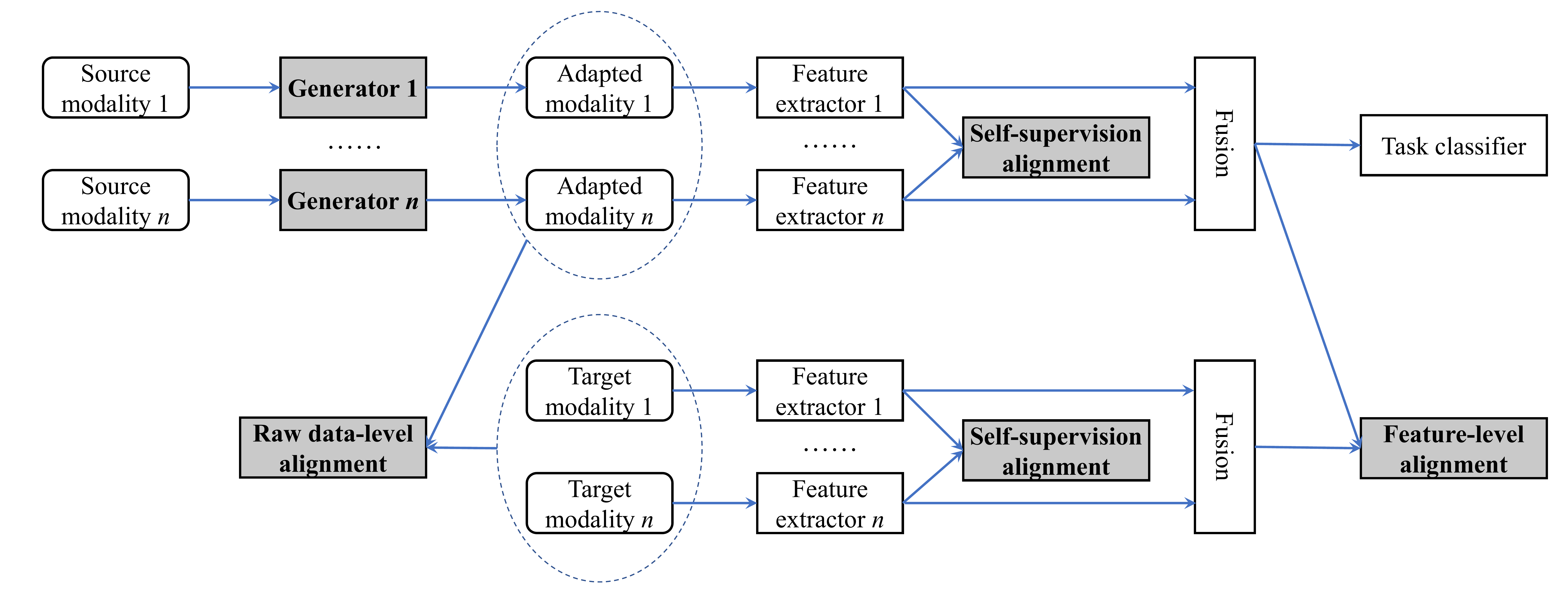}
\caption{A generalized framework for multi-modal domain adaptation with one labeled source domain and one unlabeled target domain. The gray-scale rectangles with text in bold represent different alignment strategies. Most existing multi-modal domain adaptation methods can be obtained by employing different component details, enforcing some constraints, or slightly changing the architecture.}
\label{fig:domainAdaptation}
\end{center}
\end{figure*}


\subsubsection{Quantitative Comparison of Representative MER Methods}
To give readers an impression on the performances of state-of-the-art MER methods, we conduct experiments to fairly compare some representative methods based on the released codes of CMU-Multimodal SDK\footnote{\href{https://github.com/A2Zadeh/CMU-MultimodalSDK}{https://github.com/A2Zadeh/CMU-MultimodalSDK}} and Multimodal Adaptation Gate\footnote{\href{https://github.com/WasifurRahman/BERT_multimodal_transformer}{https://github.com/WasifurRahman/MAG}}. Specifically, the compared non-deep methods include SVM, RF, and Tri-modal HMM (THMM); the compared deep methods include Multi-View LSTM (MT-LSTM), Bi-Directional Contextual LSTM (BC-LSTM), Tensor Fusion Network (TFN), Multi-attention Recurrent Network (MARN), Memory Fusion Network (MFN), fine-tuning (FT), and Multi-modal Adaptation Gate (MAG). We conduct experiments on five datasets: CMU-MOSI, YouTube, ICT-MMMO, MOUD, and IEMOCAP. All the datasets contain three modalities: face, speech, and transcript text. For visual features, Facet is used to extract per-frame basic and advanced emotions and facial action units as indicators of facial muscle movement. For acoustic features, COVAREP is employed to extract 12 Mel-frequency cepstral coefficients (MFCCs), pitch tracking and voiced/unvoiced segmenting features, glottal source parameters, peak slope parameters, and maxima dispersion quotients. For linguistic features, three different pre-trained word embeddings, i.e., GLOVE, BERT, and XLNet, are employed to obtain the word vector. The input to the non-deep methods is the early fusion of multi-modal features. For emotion classification, we use accuracy (A) and F1 as metrics; for emotion regression, we use mean absolute error (M) and the Pearson correlation (C) as metrics. Higher values indicate better performance for all the metrics, except M where lower values denote better performance.

From the results in Table~\ref{tab:GLOVE} and Table~\ref{tab:CMU-MOSI}, we have the following observations: First, the performances of deep models are generally better than non-deep ones. Second, for different datasets, the methods with the best performances are different. For example, RF achieves the best performance among non-deep models except CMU-MOSI, which demonstrates its good generalization ability, while the performance of SVM is much better than RF and THMM on CMU-MOSI.
Third, multi-class classification is more difficult than binary classification, such as 77.1 vs. 34.7 of MARN on CMU-MOSI. Fourth, comparing the same method in the two tables on CMU-MOSI, we can conclude that BERT and XLNet can provide better word embeddings than GLOVE and XLNet is generally better than BERT. Finally, although XLNet-based MAG achieves near-human level
performance on CMU-MOSI, there is still some gap and more efforts are expected to achieve even better performance than humans.

\subsection{Domain Adaptation for Multi-modal Emotion Recognition}

Domain adaptation aims to learn a transferable MER model from labeled source domains that can perform well on unlabeled target domains~\cite{zhao2020review}. Recent efforts have been dedicated to deep unsupervised domain adaptation~\cite{zhao2020review}, which employs a two-streams architecture. One stream is used to train an MER model on the labeled source domains, while the other is used to align the source and target domains. Based on the alignment strategy, existing uni-modal domain adaptation methods can be classified into different categories~\cite{zhao2020review}, such as discrepancy-based, adversarial discriminative, adversarial generative, and self-supervision-based methods.

\textit{Discrepancy-based methods} employ some distance metrics to explicitly measure the discrepancy between the source and target domains on corresponding activation layers of the two network streams. Commonly used discrepancy loss include maximum mean discrepancy, correlation alignment, geodesic distance, central moment discrepancy, Wasserstein discrepancy, contrastive Domain discrepancy, and higher-order moment matching. Besides the used discrepancy loss, there are some other differences between existing methods, such as whether the loss is domain-level or class-level, which layer the loss is operated on, whether the backbone networks share weights or not, and whether the aligned distribution is marginal or joint. \textit{Adversarial discriminative models} usually align the source and target domains with a domain discriminator by adversarially making different domains indistinguishable. The input to the discriminator ranges from original data to extracted features and the adversarial alignment can be global or class-wise. We can also consider using shared or unshared feature extractors.  \textit{Adversarial generative models} usually employ a generator to generate fake source or target data to make the domain discriminator indistinguishable from the generated and real domains. The generator is typically based on generative adversarial network (GAN) and its variants, such as CoGAN, SimGAN, and CycleGAN. The input to the generator and discriminator can be different in different methods. \textit{Self-supervision-based methods} combine some auxiliary self-supervised learning tasks, such as reconstruction, image rotation prediction, jigsaw prediction, and masking, with the original task network to bring the source and target domains closer. We can compare these four types of domain adaptation methods from the perspectives of theory guarantee, efficiency, task scalability, data scalability, data dependency, optimizability, and performance. We can combine some of these methods to jointly exploit their advantages.

The main difficulty in domain adaptation for MER lies in the alignment of multiple modalities between the source and target domains simultaneously. There are some simple but effective ways to extend uni-modal domain adaptation to multi-modal settings, as shown in Fig.~\ref{fig:domainAdaptation}. For example, we can use discrepancy loss or discriminator to align the fused feature representations. The correspondence between different modalities can be used as a self-supervised alignment. Extending adversarial generative models from uni-modal to multi-modal would be more difficult. Unlike image, other generated modalities, such as text and speech, might have confused semantics, although they can make the discriminator indistinguishable. Generating intermediate feature representations instead of raw data can provide a feasible solution.

\section{Applications}
Recognizing emotions from multiple explicit cues and implicit stimuli is of great significance in a broad range of real-world applications.
Generally speaking, emotion is the most important aspect of the quality and meaning of our existence, which makes life worth living.
The emotional impact of digital data lies in that it can improve the user experience of existing techniques and then strengthen the knowledge transfer between people and computers~\cite{joshi2011aesthetics}.
Many people tend to post texts, images, and videos on social networks to express their daily life feelings.
Inspired by this, we can mine people's opinions and sentiments towards topics and events happening in the real world~\cite{ji2019weibo}.
For instance, user-generated content in Facebook or Instagram can be used to derive the attitudes of people from different countries and regions when they face epidemics like COVID-19~\cite{lyu2020covid}.
Researchers also try to detect sentiment in social networks and apply the results to predict political elections.
Note that when the personalized emotion of an individual is detected, we can further group these emotions, which may contribute to predicting the tendencies of society.
Another important application of multi-modal emotion recognition is business intelligence, especially marketing and consumer behavior analysis~\cite{wu2017regret}.
Nowadays, most apparel e-retailers use human models to present products.
The model's face presentation is proved to have a significant effect on consumer approach behavior.
To be specific, for participants whose emotional receptivity is high, smiling facial expression tends to lead to the highest approach behavior.
Besides, researchers examine how online store specialization influences consumer pleasure and arousal, based on the stimulus-organism-response framework.
Emotion recognition can also be used in call centers, the goal of which is to detect the emotional states of both the caller and the operator.
The system recognizes the involved emotions through the intonation and tempo, as well as the texts translated from the corresponding speech.
Based on this, we can receive feedback on the quality of the service.

Meanwhile, emotion recognition plays an important role in the field of medical treatment and psychological health.
With the popularity of social media, some people prefer to sharing their emotions on the Internet rather than with others.
If a user is observed to be sharing negative information (e.g., sadness) frequently and continuously, it is necessary to track her/his mental status to prevent the occurrence of psychological illness and even suicide.
Emotional states can also be used to monitor and predict fatigue states of a variety of people like drivers, pilots, workers in assembly lines, and students in classrooms.
This technique both prevents dangerous situations and benefits the evaluation of work/study efficiency.
Further, emotional states can be incorporated into various security applications, such as systems for monitoring public spaces (e.g., bus/train/subway stations, football stadiums) for potential aggression.
Recently, an effective auxiliary system is introduced in the diagnosis and treatment process of autism spectrum disorder (ASD) of children, to assist in collecting the pathological information.
To help professional clinicians better and faster make a diagnosis and give treatment to ASD patients, this system characterizes facial expressions and eye gaze attention which are considered to be remarkable indicators for early screening of autism.
%

%
Multi-modal emotion recognition is used to improve the personal entertainment experience.
For example, a recent work in brainwave–music interface maps EEG characteristics to musical structures (note, intensity, and pitch).
Similarly, efforts have been made to understand the emotion-centric correlation between different modalities that are essential for various applications.
Affective image-music matching provides a good chance to append a sequence of music to a given image, where they may evoke the same emotion.
This helps generate emotion-aware music playlists from one's personal album photos in mobile devices.

\section{Future directions}
Existing methods have achieved promising performances on various MER settings, such as visual-audio, facial-textual-speech, and textual-visual tasks. However, all the summarized challenges have not been fully addressed. For example, how to extract discriminative features that are more related to emotion, how to balance between common and personalized emotion reactions, and how to emphasize the more important modalities are still open. To help improve the performances of MER methods and make them fit special requirements in real world, we provide some potential future directions.

\textbf{New Methodologies for MER.} \textit{1) Contextual and prior knowledge modeling.} The experienced emotion of a user can be significantly influenced by the contextual information, such as the conversational and social environments. The prior knowledge of users, such as personality and age, can also contribute to emotion perception. For example, an optimistic user and a pessimistic viewer are likely to see different aspects of the same stimuli. Jointly considering these important contextual and prior knowledge is expected to improve the MER performance. Graph-related methods, such as graph convolutional networks, are possible solutions to model the relationships between factors and emotions.
\textit{2) Learning from unlabeled, unreliable, and unmatched affective signals.} In the big data era, the affective data might be sparsely labeled or even unlabeled, the raw data or labels can be reliable, and the test and training data might be unmatched. Exploring advanced machine learning techniques, such as unsupervised representation learning, dynamic data selection and balancing, and domain adaptation, and embedding the special properties of emotions, can help to address these challenges. \textit{3) Explainable, robust, and secure deep learning for MER.} Due to the black-box nature, it is difficult to understand why existing deep neural networks perform well for MER and the trained deep networks are vulnerable to adversarial attacks and inevitable noises that might cause erraticism. Essentially explaining the decision-making process of deep learning can help to design robust and secure MER systems. \textit{4) Combination of explicit and implicit signals.} Both explicit and implicit signals are demonstrated to be useful for MER but they also suffer from some limitations. For example, explicit signals can be easily suppressed or are difficult to capture, while implicit signals might not reflect the emotions in real-time.
Jointly combining them to explore the complementary information during viewer-multimedia interaction would boost the MER performance. \textit{5) Incorporation of emotion theory into MER.} Different theories have been proposed in psychology, physiology, neurology, and cognitive sciences. These theories can help to understand how humans produce emotion but have not been employed in the computational MER task. We believe such incorporation would make more sense to recognize emotions.

\textbf{More Practical MER Settings.} \textit{1) MER in the wild.} Current MER methods mainly focus on neat lab settings. However, MER problems in the real world are much more complex. For example, the collected data might contain much noise that is unrelated to emotion; the users in the test set are from different
cultures and languages from those in the training set, which results in different ways of emotion expression; different emotion label spaces are employed across various settings; training data is incrementally available. Designing an effective MER model that is generalizable to these practical settings is worth investigating. \textit{2) MER on the edge.} When deploying MER models in edge devices, such as mobile phones and security cameras, we have to consider the computing limitation and data privacy. Techniques like auto pruning, neural architecture search, invertible neural network, and software-hardware co-design are believed to be beneficial for efficient on-device training. \textit{3) Personalized and group MER.} Because of the emotion's subjectivity, simply recognizing the dominant emotion of different individuals is insufficient. It is ideal but impractical to collect enough data for each individual to train personalized MER models. Adapting the well-trained MER models for dominant emotions to each individual with a small amount of labeled data is a possible alternate solution. On the other hand, it would make more sense to predict emotions for groups of individuals who share similar tastes or interests and have a similar background. Group emotion recognition is essential in many applications, such as recommendation systems, but how to classify users into different groups is still challenging.

\textbf{Real Applications Based on MER.} \textit{1) Implementation of MER in real-world applications.} Although emotion recognition has been emphasized to be important for decades, it has rarely been applied to real scenarios due to relatively low performance. With the recent rapid progress of MER, we can begin incorporating emotion into different applications in marketing, education, health care, and service sectors. The feedback from the applications can in turn promote the development of MER. Together with emotion generation, we believe an age of artificial emotional intelligence is coming. \textit{2) Wearable, simple, and accurate affective data collection.} To conduct MER tasks, the first step is to collect accurate affective data. Developing wearable, simple and even contactless sensors to capture such data would make users more acceptable. \textit{3) Security, privacy, ethics, and fairness of MER.} During data collection, it is possible to extract users' confidential information, such as identity, age, etc. Protecting the security and privacy of users and avoiding any chance of misuse must be taken into consideration. Emotion recognition in real applications might have a negative and even dangerous impact on a person, such as emotional pressure. Methods to eliminate such impact should also be considered from the perspectives of ethics and fairness.

\section{Conclusion}
In this article, we provided a comprehensive tutorial on multi-modal emotion recognition (MER). We briefly introduced emotion representation models, both explicit and implicit affective modalities, emotion annotations, and corresponding computational tasks. We summarized the main challenges of MER in detail, and then we emphatically introduced different computational methodologies, including representation learning of each affective modality, feature fusion of different affective modalities, classifier optimization for MER, and domain adaptation for MER. We ended this tutorial with discussions on real-world applications and future directions. We hope this tutorial can motivate novel techniques to facilitate the development of MER, and we believe that MER will continue to attract significant research efforts.

\noindent{\textbf{Acknowledgements}: This work was supported by the National Key Research and Development Program of China Grant (No. 2018AAA0100403), the National Natural Science Foundation of China (Nos. 61701273, 61876094, U1933114, 61925107, U1936202), the Natural Science Foundation of Tianjin, China (Nos.20JCJQJC00020, 18JCYBJC15400, 18ZXZNGX00110), and Berkeley DeepDrive.}

\ifCLASSOPTIONcaptionsoff
  \newpage
\fi


\bibliographystyle{IEEEtranN}
\footnotesize{\bibliography{IEEESPM}}

\begin{IEEEbiographynophoto}{Sicheng Zhao} (SM'19) received the Ph.D. degree from Harbin Institute of Technology, Harbin, China, in 2016. He was a Visiting Scholar at National University of Singapore from July 2013 to June 2014, a Research Fellow at Tsinghua University from September 2016 to September 2017, and a Research Fellow at University of California, Berkeley from September 2017 to September 2020. He is currently a Postdoc Research Scientist at Columbia University, USA. His research interests include affective computing, multimedia, and computer vision.
\end{IEEEbiographynophoto}

\vspace{-30pt}

\begin{IEEEbiographynophoto}{Guoli Jia}
will work toward the Master’s degree at the College of Computer Science, Nankai University, Tianjin, China. His research interests include computer vision and pattern recognition.
\end{IEEEbiographynophoto}

\vspace{-30pt}

\begin{IEEEbiographynophoto}{Jufeng Yang} received the Ph.D.\ degree from Nankai University, Tianjin, China, in 2009. He is currently a full professor in the College of Computer Science, Nankai University and was a visiting scholar with the Vision and Learning Lab, University of California, Merced, USA, from 2015 to 2016. His research falls in the field of computer vision, machine learning and multimedia. His recent interests include affective computing, image retrieval, fine-grained classification, and medical image recognition.
\end{IEEEbiographynophoto}

\vspace{-30pt}

\begin{IEEEbiographynophoto}{Guiguang Ding} received his Ph.D.\ degree from Xidian University, China, in 2004. He is currently a full professor with School of Software, Tsinghua University. Before joining school of software in 2006, he has been a postdoctoral research fellow in the Department of Automation, Tsinghua University. His current research centers on the area of multimedia information retrieval, computer vision, and machine learning. He served as a leading guest editor of NPL and MTAP, a special session chair of ICASSP 2021, ICME 2020/2019, PCM 2017, and a reviewer for over 20 prestigious international journals and conferences. He has published over 100 scientific papers in major journals and conferences.
\end{IEEEbiographynophoto}

\vspace{-30pt}

\begin{IEEEbiographynophoto}{Kurt Keutzer} received his Ph.D.\ degree in Computer Science from Indiana University in 1984 and then joined the research division of AT\&T Bell Laboratories. In 1991 he joined Synopsys, Inc.\ where he ultimately became Chief Technical Officer and Senior Vice-President of Research. In 1998, Kurt became Professor of Electrical Engineering and Computer Science at the University of California at Berkeley. Kurt's research group is currently focused on using parallelism to accelerate the training and deployment of Deep Neural Networks for applications in computer vision, speech recognition, multi-media analysis, and computational finance. Kurt has published six books, over 250 refereed articles, and is among the most highly cited authors in Hardware and Design Automation. Kurt is a Life Fellow of the IEEE.
\end{IEEEbiographynophoto}

\end{document}